\documentclass[11pt,dvips]{article}

\usepackage{epsfig,times}
\usepackage{picinpar}
\usepackage{wrapfig}
\usepackage{floatflt}
\makeatletter
\renewcommand{\section}{\@startsection{section}{1}{0in}
    {0.4\baselineskip}{0.1\baselineskip}{\Large\bf}}
\renewcommand{\subsection}{\@startsection{subsection}{2}{0in}
    {0.25\baselineskip}{-\baselineskip}{\large\bf}}
\renewcommand{\subsubsection}{\@startsection{subsubsection}{3}{0in}
    {0.1\baselineskip}{-\baselineskip}{\normalsize\bf}}
\makeatother
\begin{document}
\makeatletter\newcommand{\ps@icrc}{
\renewcommand{\@oddhead}{\slshape{OG.2.3.14}\hfil}}
\makeatother\thispagestyle{icrc}
\begin{center}
\begin{center}
%
{\LARGE \bf Gamma Ray Burst and Soft Gamma Repeaters.\\
Spinning,precessing  $\gamma$ jets}

\end{center}


%

\begin{center}
{\bf D. Fargion $^1$}\\ {\it $^{1}$ Physics Department, Rome
University 1, and INFN, Rome1, P.za Aldo Moro 2 Rome, ITALY}
\end{center}
\begin{center}
{\large \bf Abstract\\}
\end{center}
\end{center}

\vspace{-0.5ex}
%
%
Gamma Ray Bursts as recent GRB990123 and  GRB990510 are observed
to occur in cosmic volumes with a corresponding output
reaching,for isotropic explosions, energies as large as  two solar
masses annihilation. These energies are underestimated because of
the neglected role of comparable ejected neutrinos bursts. These
extreme power cannot be explained with any standard spherically
symmetric Fireball model. A too heavy black hole or Star would be
unable to coexist with the shortest millisecond time structure of
Gamma ray Burst. Beaming of the gamma radiation may overcome the
energy puzzle. However any mild explosive beam $(\Omega > 10^{-2}
)$ should not solve the jet containment at those disruptive
energies. Only extreme beaming $(\Omega < 10^{-8} )$, by a slow
decaying, but long-lived precessing jet, it may coexist with
characteristic Supernova energies, apparent GRBs output,
statistics as well as their connection with older and nearer SGRs
relics.
%

\vspace{1ex}

%
%

\section{Introduction}
After a decade, at present (GRB990123 over energetic event) none
spent a regret word on the decline and possible final rejection of
the popular isotropic burst fireball model. GRBs connections with
jets is growing from supernova connections, energy crisis and
polarization evidences. On the other hand SGRs have still a
popular magnetar (isotropic mini-fireball) model well alive. It is
therefore time to remind that recent strong SGR events on 1998
(SGR1900+14), (SGR1642-21),as well as the old 5th March79 SGR,
shared the same hard spectra of classical GRBs. It is in
particular very instructive to notice the GRB-SGR similar spectra
morphology and temporal evolution within BATSE trigger 7172
GRB981022 and 7171 GRB981022. Nature would be extremely perverse
to mimic two very similar events (either for time structures and
energy spectra) at same detector and at the same day  by two
totally different processes. A magnetar mini-fireball (for SGRs)
versus the GRB burst, at the present more related to jets. We
argue here that, apart of the energetics, both of them are blazing
of powerful jets (NS or BH) by spinning and precessing source in
either binary or in accreting disk systems (Fargion 1998). The
GRBs optical transient after-glows are the Supernova like
explosive birth of the jet. Their optical flash,days after the
burst, is related to the maximal optical explosion intensity and
it is enhanced only by a partial beaming $(\Omega \simeq 10^{-2}
)$. The rarest extreme peak OT during GRB990123 (at a million time
a Supernova luminosity)  is the beamed $(\Omega \leq 10^{-5} )$
Inverse Compton optical tail responsible of the same extreme gamma
(MeV) extreme beamed $(\Omega \leq 10^{-8})$ signal. The huge
energy bath (for a fireball model) on GRB990123 imply the
coexistence of an energetic neutrino burst comparable to the
photon one. Indeed , in analogy to the early three minutes of the
hot universe, if entropy conservation holds, the energy density
factor to be added to the photon $\gamma$ GRB990123 budget is at
least $( \simeq (21/8)\times (4 /11)^{4/3} )$. In this case the
final gamma energy enjoy of the electron pairs annihilation, as in
the thermal equilibrium in the first second of the universe. If
the GRB had not time to keep the entropy conservation (the most
probable case) than the energy needed for the neutrino burst was
at least a factor $[21/8]$ larger than the gamma one. The
consequent energy-mass needed for the two cases (including both
$\nu$ and $\gamma$ burst) are respectively 3.5 and 7.2 solar
masses. No known isotropic fireball model may release at ideal
total energy conversion  such a huge energy burst. One must also
remind that maximal black hole energy conversion takes place for
rotational case at a level below  a factor $0.4$. Therefore the
original masses for isotropic fireball must require at least a
$8,7$ solar mass black hole, with obvious contradictions with
millisecond gamma burst fine structures.
 Beaming may solve the puzzle. Extreme $\nu$ and $\gamma$
 beaming by a rapid spinning and precessing jet,
 (a neutron star or a black hole), may explain the
apparent extreme energy. Also the over supernova optical transient
peak intensities are beamed within a thin jet.
 We therefore predict here that future detailed
(within fraction of second detection) observations of this
contemporaneous (seconds delay) optical transient events must be
modulated in a fine structured way,  nearly comparable to the
gamma ray burst signal.
 Moreover the GRB980425-SN1998bw (Galama et
al. 1998) association put already since a year in severe strain
any ``candle'' fireball. Indeed isotropic standard candle
(luminosity $l_\gamma$) fireballs are unable to explain the
following key questions related to that GRB-SN association:
\begin{enumerate}
 \item Why nearest ``local'' GRB980425 in ESO 184-G82 galaxy at
 redshift $z_2 = 0.0083$ and the most far away ``cosmic'' ones as
 GRB971214 (Kulkarni et al.1998) at redshift $z_2 = 3.42$
 exhibit a huge average and peak
 intrinsic luminosity ratio?
\begin{equation}\label{eq1}
\frac{<L_{1 \gamma}>}{<L_{2 \gamma}>}  \cong  \frac{<l_{1
\gamma}>}{<l_{2 \gamma}>} \frac{z_{1 }^2}{z_{2 }^2} \cong 2 \cdot
10^5 \;\;; \left. \frac{L_{1 \gamma}}{L_{2 \gamma}} \right|_{peak}
\simeq 10^7 .
\end{equation}
Fluence ratios $E_1 / E_2$ are also extreme ($\geq 4 \cdot 10^5$).
 \item Why GRB980425 nearest event spectrum is softer than cosmic
 GRB971214 while Hubble expansion would imply the opposite by a
 redshift factor $(1+z_1)\sim 4.43$?
 \item Why,  GRB980425 time structure is
 slower and smoother than cosmic one,as above contrary to Hubble law?
 \item Why we observed so many (even just the rare April one over 14
Beppo Sax optical transient event) nearby GRBs? Their probability
to occur, with respect to a cosmic redshift  $z_1 \sim 3.42$ must
be suppressed by a severe volume factor
\begin{equation}\label{eq2}
\frac{P_{1}}{P_{2}} \cong \frac{z_{1}^{3}}{z_{2}^{3}} \simeq 7
\cdot 10^{7} \;\;\;.
\end{equation}
\end{enumerate}
The above questions remain unanswered by fireball candle model. A
family of new GRB fireballs are ad hoc and fine-tuned solutions.
We believed since 1993 (Fargion 1994) that spectral and time
evolution of GRB are made up blazing beam gamma jet GJ. The GJ is
born by ICS of ultrarelativistic (1 GeV-tens GeV) electrons
(pairs) on source IR, or diffused companion IR, BBR photons
(Fargion,Salis 1998). The beamed electron jet pairs will produce a
coaxial gamma jet. The simplest solution to solve the GRBs
energetic crisis (as GRB990123 whose isotropic budget requires an
energy above two solar masses) finds solution in a geometrical
enhancement by the jet thin beam. A jet angle related by a
relativistic kinematics would imply $\theta \sim
\frac{1}{\gamma_e}$, where $\gamma_e$ is found to reach $\gamma_e
\simeq 10^3 \div 10^4$ (Fargion 1994,1998). However an impulsive
unique GRB jet burst (Wang \& Wheeler 1998) increases the apparent
luminosity by $\frac{4 \pi}{\theta^2} \sim 10^7 \div 10^9$ but
face a severe probability puzzle due to the rarity to observe a SN
burst jet pointing in line toward us. Therefore we considered
 GRBs and SGRs as multiprecessing
and spinning Gamma Jets. In particular we considered (Fargion
1998) an unique scenario where primordial GRB jets decaying in
hundred and thousand years become the observable nearby SGRs. The
ICS for monochromatic electrons on  BBR leads to a coaxial gamma
jet spectrum(Fargion \& Salis 1995,1996,1998):
$\frac{dN_{1}}{dt_{1}\,d\epsilon_{1}\,d\Omega _{1}}$ is
\begin{equation}
\epsilon _{1}\ln \left[ \frac{1-\exp \left( \frac{-\epsilon
_{1}(1-\beta \cos \theta _{1})}{k_{B}\,T\,(1-\beta )}\right)
}{1-\exp \left( \frac{-\epsilon _{1}(1-\beta \cos \theta
_{1})}{k_{B}\,T\,(1+\beta )}\right) }\right] \left[ 1+\left(
\frac{\cos \theta _{1}-\beta }{1-\beta \cos \theta _{1}}\right)
^{2}\right] \label{eq3}
\end{equation}
scaled by a proportional factor $A_1$ related to the electron jet
intensity. The adimensional photon number rate (Fargion \& Salis
1996) as a function of the observational angle $\theta_1$
responsible for peak luminosity (eq. \ref{eq1}) becomes
\begin{equation}
\frac{\left( \frac{dN_{1}}{dt_{1}\,d\theta _{1}}\right) _{\theta
_{1}(t)}}{ \left( \frac{dN_{1}}{dt_{1}\,d\theta _{1}}\right)
_{\theta _{1}=0}}\simeq \frac{1+\gamma ^{4}\,\theta
_{1}^{4}(t)}{[1+\gamma ^{2}\,\theta _{1}^{2}(t)]^{4}}\,\theta
_{1}\approx \frac{1}{(\theta _{1})^{3}} \;\;.\label{eq4}
\end{equation}
The total fluence at minimal impact angle $\theta_{1 m}$
responsible for the average luminosity (eq. \ref{eq1}) is
\begin{equation}
\frac{dN_{1}}{dt_{1}}(\theta _{1m})\simeq \int_{\theta
_{1m}}^{\infty }\frac{ 1+\gamma ^{4}\,\theta _{1}^{4}}{[1+\gamma
^{2}\,\theta _{1}^{2}]^{4}} \,\theta _{1}\,d\theta _{1}\simeq
\frac{1}{(\,\theta _{1m})^{2}}\;\;\;. \label{eq5}
\end{equation}
These spectra fit GRBs observed ones (Fargion \& Salis 1995).
Assuming a beam jet intensity $I_1$ comparable with maximal SN
luminosity, $I_1 \simeq 10^{45}\;erg\,s^{-1}$, and replacing this
value in adimensional $A_1$ in equation \ref{eq3} we find a
maximal apparent GRB power for beaming angles $10^{-3} \div
3\times 10^{-5}$, $P \simeq 4 \pi I_1 \theta^{-2} \simeq 10^{52}
\div 10^{55} erg \,s^{-1}$ within observed ones. We also assume a
power law jet time decay as follows
\begin{equation}\label{eq6}
  I_{jet} = I_1 \left(\frac{t}{t_0} \right)^{-\alpha} \simeq
  10^{45} \left(\frac{t}{3 \cdot 10^4 s} \right)^{-1} \; erg \,
  s^{-1}
\end{equation}
where ($\alpha \simeq 1$) able to reach, at 1000 years time
scales, the present known galactic microjet (as SS433) intensities
powers: $I_{jet} \simeq 10^{38}\;erg\,s^{-1}$. We used the model
to evaluate if April precessing jet might hit us once again.
\section{The GRB980425-GRB980712 repeater}
Therefore the key answers to the puzzles (1-4)  are: the GRB980425
has been observed off-axis by a cone angle wider than
$\frac{1}{\gamma}$ thin jet  by a factor $a_2 \sim 500$ (Fargion
1998) and therefore one observed only the ``softer'' cone jet tail
whose spectrum is softer and whose time structure is slower
(larger impact parameter angle). A simple statistics favoured a
repeater hit. Indeed GRB980430 trigger 6715 was within $4 \sigma$
and particularly in GRB980712 trigger 6917 was within $1.6 \sigma$
angle away from the April event direction. An additional event 15
hours later, trigger 6918, repeated making the combined
probability to occur quite rare ($\leq 10^{-3}$). Because the July
event has been sharper in times ($\sim 4 \,s $) than the April one
($\sim 20 \,s $), the July impact angle had a smaller factor $a_3
\simeq 100$. This value is well compatible with the
 expected peak-average luminosity flux evolution in eq.(6,4):
$\frac{L_{04\,\gamma}}{L_{07\,\gamma}} \simeq
\frac{I_2\,\theta_2^{-3}}{ I_3\,\theta_3^{-3}} \simeq \left(
\frac{t_3}{t_2} \right)^{-\alpha} \,\left( \frac{a_2}{a_3}
\right)^{\,3} \leq 3.5$ where $t_3 \sim 78 \; day$ while $t_2 \sim
2 \cdot 10^5 \, s$. The predicted fluence is also comparable with
the observed ones $\frac{N_{04}}{N_{07}} \simeq
\frac{<L_{04\,\gamma}>}{<L_{07\,\gamma}>} \, \frac{\Delta
\tau_{04}}{\Delta \tau_{07}} \simeq \left( \frac{t_3}{t_2}
\right)^{-\alpha} \,\left( \frac{a_2}{a_3} \right)^2
\,\frac{\Delta \tau_{04} }{\Delta \tau_{07}} \geq 3$.
\section{The SGRs hard spectra and their GRB link}
Last SGR1900+14 (May-August 1998) events and SGR1627-41
(June-October 1998) events did exhibit at peak intensities hard
spectra comparable with classical GRBs. We imagine their nature as
the late stages of jets fueled by a disk or a companion (WD,NS)
star. Their binary angular velocity $\omega_b$ reflects the beam
evolution $\theta_1(t) = \sqrt{\theta_{1 m}^2 + (\omega_b t)^2}$
or more generally a multiprecessing angle $\theta_1(t)$ (Fargion
\& Salis 1996) wich keeps memory of the pulsar jet spin
($\omega_{psr}$), precession by the binary $\omega_b$ and
additional nutation due to inertial momentum anisotropies or
beam-accretion disk torques ($\omega_N$). On average, from eq.(5)
the gamma and afterglow decays as $t^{-2}$; the complicated
spinning and precessing jet blazing is responsible for the wide
morphology of GRBs and SGRs as well as their internal periodicity.
In conclusion the puzzles for GRB980425-GRB971214 find a simple
solution within a precessing jet: the different geometrical
observational angle may compensate the April 1998 low peak gamma
luminosity ($10^{-7}$) by a larger impact angle which compensates,
at the same time, the statistical rarity ($\sim 10^{-7}$) to find
in a near volume a GRBs, its puzzling softer nature as well as its
longer (larger impact parameter view) timescales. Finally the
April GRB repetitivity on GRB980712 verified the model. Such
precessing jets may also explain (Fargion \& Salis 1995) the
external twin rings around SN1987A. They may propel and inflate
plerions as the observed ones near SRG1647-21 and SRG1806-20. In
conclusion optical nebula NGC6543 (``Cat Eye'') and its thin jets
fingers as well as the inexplicable double cones sections in Egg
nebula CRL2688 are the  spectacular lateral view of such spinning
and precessing jets. Their blazing in-axis toward us  would appear
as SGRs . At their  maximal power during their SN birth, their
blazing would appear  as a GRBs marked by their  coeval optical
afterglow.

%
%
%
%
%
%
\vspace{1ex}
\begin{center}
{\Large\bf References}
\end{center}
%
 Fargion, D.: 1994, The Dark Side of the Universe. R. Bernabei,June
 1993, World Scientific, p.88-97
 \\
 Fargion, D., Salis, A.: 1995, Nuclear Phys B (Proc. Suppl.) 43, 269-273
 \\

%
 Fargion, D., Salis, A.: 1995, NATO ASI, 461, 397-408
 \\
 Fargion, D., Salis, A.: 1996, 3rd GRB: AIP. Conf. 384; 754-758
 \\
 Fargion, D., Salis, A.: 1996, 3rd GRB: AIP. Conf. 384; 749-753
 \\
 Fargion, D.: 1998a, The Astronomers Telegram. Atel \# 31
 \\
 Fargion, D.: 1998b, astro-ph/9808005
 \\
 Fargion, D., Salis, A.: 1998, Physics-Uspekhi, 41(8), 823-829
\\
 Galama, T. J., et al.: 1998, Astro-ph/9806175
 \\
 Kulkarni, S. R., et al., 1998, Nature 393, 35-39
 \\
 Wang L., Wheeler, J. C.: 1998,Astro-ph/9806212
 \\
\end{document}